\documentclass[aps,prb,twocolumn,groupedaddress]{revtex4}
\usepackage{graphics}
\usepackage{epsfig}
\usepackage{amsmath}
\usepackage{amssymb}
\newcommand{\ds}{\displaystyle}

\begin{document}

\title{Diffusion-limited unbinding of small peptides from PDZ domains}

\author{Fabio Cecconi$^1$,  Paolo De Los Rios$^2$ and  Francesco Piazza$^2$} 

\affiliation{
        $^1$INFM-SMC and Istituto dei Sistemi Complessi ISC-CNR 
              Via dei Taurini 19, 00185 Roma, Italy\\                                                   
       $^2$Laboratoire de Biophysique Statistique, 
          EPFL SB ITP, CH-1015, Lausanne, Switzerland.}

\begin{abstract}
PDZ domains are typical examples of binding motifs 
mediating the formation of protein-protein assemblies in many different
cells. A quantitative characterization of the mechanisms intertwining structure, 
chemistry and dynamics with the PDZ function represent a challenge in molecular 
biology. Here we investigated the influence of native state topology 
on the thermodynamics and  the dissociation kinetics for a complex 
PDZ-peptide via Molecular Dynamics simulations
based on a coarse-grained description of PDZ domains. 
Our native-centric approach neglects chemical details but incorporates the basic 
structural information to reproduce the protein functional dynamics as it couples
to the binding. We found that at physiological temperatures the unbinding of a peptide 
from the PDZ  domain becomes increasingly diffusive rather than thermally activated, 
as a consequence of the significant reduction of the free energy barrier with temperature.
In turn, this results in a significant slowing down of the process of two
orders of magnitude with respect to the conventional Arrhenius extrapolation from low 
temperature calculations.
Finally, a detailed analysis of a typical unbinding event based on the rupture times 
of single peptide-PDZ contacts allows to shed further light on the dissociation 
mechanism and to elaborate a coherent picture of the relation between function
and dynamics in PDZ domains.
\end{abstract}

\maketitle

\section{Introduction}

The role of PDZ domains in the organization of protein complexes at 
the plasma membrane has been increasingly recognized in the last
decade~\cite{Fanning:1999}. Proteins containing several PDZ domains (up to 13
in the MUPP1 protein~\cite{Ullmer:1998}) act as scaffolds that cluster together
different transmembrane, membrane associated and periplasmic proteins involved,
among other functions, in signaling pathways ~\cite{Pawson:1997,Fan:2002} and
ion permeability~\cite{Bezprozvanny:2001a}. The participation of PDZ domains in
the organization of supramolecular complexes in skeletal muscle cells has also
been documented~\cite{Kachinsky:1999,Faulkner:1999}.

PDZ domains associate with other proteins by binding their carboxyl-terminal
aminoacids~\cite{Sheng:2001,Hung:2002}, as highlighted by the structure of
several PDZ domains, such as the third PDZ domain of PSD95 (Postsynaptic
density-95/disks large/zonula occludens-1)~\cite{Doyle:1996} and the PDZ
domain of ZASP (Z-band alternatively spliced PDZ-motif) ~\cite{Au:2004},
although internal structures, such as $\beta$-hairpins that mimic C-terminal
geometries can also be recognized, as in the complex between nNOS (neuronal
nitric oxide synthase) and syntrophin~\cite{Hillier:1999}.

Recently, due to their central role as key mediators of 
protein--protein interactions in mammalian cells, PDZ domains have 
been the object of intense study, with the aim 
of designing small molecules capable of acting as modulators or 
inhibitors of the PDZ binding activity in a controlled fashion. 
Efforts have focused in the design of both 
non--peptide~\cite{Fujii:2002nx,Fujii:2003lq,Shan:2005dq} and 
peptide~\cite{Saro:2003wd,Wiedemann:2004cr,Udugamasooriya:2005rr} ligands, 
with the ambition to develop molecular probes to study the biophysical 
and biochemical properties of PDZ domains and to devise 
new small molecule--based 
therapeutic strategies. Hence, the great importance of 
understanding the principles of peptide--PDZ interactions.

\begin{figure}[t!]
\includegraphics[clip=true,keepaspectratio,width=6.0truecm]{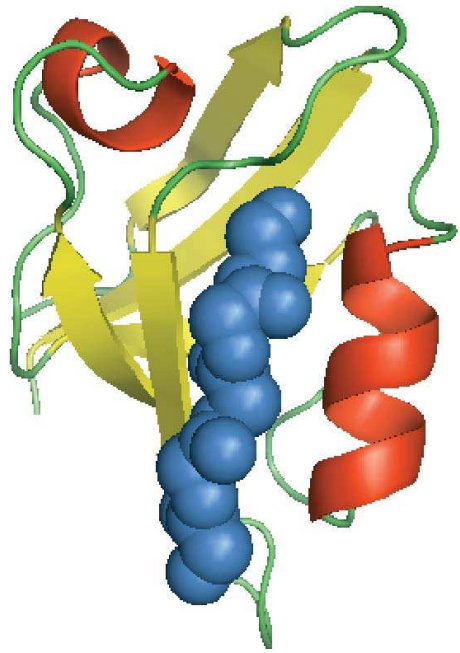}
\caption{Cartoon representation of the crystal structure of the PDZ3 domain 
in complex with its target peptide (blue). PDB code 1BFE.}
\label{fig:struct}
\end{figure}

The geometry and chemistry of binding to PDZ domains involve the fit of the
last 4 to 5 carboxyl-terminal aminoacids into a groove between a $\alpha$-helix
and a $\beta$-strand on the PDZ surface (Fig.\ref{fig:struct}), 
with the last C-terminal residue
almost invariably hydrophobic. The specificity of each domain is conferred by a
few (2 to 3) PDZ surface aminoacids that make contacts with the residues in
positions -1 to -4 relative to the C-terminal in the target
protein~\cite{Sheng:2001}. The surprising simplicity of this binding  scheme
possibly explains why PDZ domains are one of the most widespread binding
modules yet identified, since just a few incremental, concerted mutations
involving surface  aminoacids (hence unlikely to change the overall protein
stability) can tune the affinity of  PDZ domains for different targets. On the
other hand, a binding architecture that relies on just a few optimized contacts
comes at the price of losing strict specificity. Indeed, recent experiments on 26
mouse PDZ domains and domain clusters have confirmed that each PDZ domain can
bind to several peptides~\cite{Stiffler:2006}, and that each peptide, 
in turn, can bind to several PDZ domains.

In a recent work \cite{immortal}, we have explored  
via normal mode analysis~(NMA) the mechanical aspects of binding of a peptide
to a  PDZ domain. In line with the results of NMA analysis of 
several proteins~\cite{Tama:01,Delarue:02,Gerstein:98}, the picture that
emerged was that  a limited number of low-frequency modes suffice to
reconstruct the  observed conformational change from the apo to the complexed
structure of the PDZ. The long-range spatial correlations that characterize
these modes correspond to a  concerted  {\em breathing} motion of the binding
cleft, thus suggesting that  functional dynamics is deeply rooted in the
native architecture itself. 
However, despite their success in providing a qualitative
picture of the coupling between thermal fluctuations of the structure and the
binding deformation, the predictive power of normal  modes is still limited by
the harmonic approximation. Hence, in order to investigate the progressive
detachment of the peptide from the PDZ structure, one must resort to more
general computational studies.

Molecular dynamics simulations have played a crucial role in
understanding the basis of the concomitant selectivity and promiscuity of the
PDZ binding dynamics~\cite{Madsen:2005,Basdevant:2006}.  Basdevant 
{\it et al.}~\cite{Basdevant:2006}, 
in particular, have used all-atom Molecular Dynamics (MD)
with a realistic force-field to correctly reproduce 
the experimental ranking, but
not the precise values, of binding free-energies. The chemical details of the
target peptide and of the binding groove have been taken into account, and
hydrophobic effects have been shown to be key determinants of the promiscuity,
with other  interactions conferring stronger or weaker selectivity. 

The free energy difference $\Delta G_b^u$ between the unbound and bound states is
customarily related to the dissociation constant $K_D$ by  $K_D=\exp(-\Delta
G_b^u/k_B T)$ where T is the absolute temperature and $k_B$ is the Boltzmann
constant. However, the constant $K_D$ is also related to the dynamics of
the system by the equation $K_D=k_{\rm off}/k_{\rm on}$, 
where  $k_{\rm off}$ and $k_{\rm
on}$ are the unbinding and binding rates respectively. At a first
approximation, $k_{\rm off}$ and $k_{\rm on}$ depend on the free energy 
difference between the bound and unbound
states and the barrier between them. The dissociation constant $K_D$, therefore,
conceals the kinetics of the system, the same value being compatible with both
very fast and very slow rates, as long as their ratio does not change. 
In the
case of PDZ domains, dissociation rates range experimentally~\cite{Gianni:2005a,Songyang:1997} 
from  a few s$^{-1}$ to 10$^{-3}$ s$^{-1}$,
thus implying that a correct description of  the unbinding process would need
to capture the PDZ dynamics over comparable timescales.  Unfortunately,
detailed all-atom MD simulations can at present cover  a few
tens~\cite{Basdevant:2006} and at most reach a hundred nanoseconds. Therefore, in order to
observe a typical unbinding event, MD simulations should be from 6 to 10 orders
of magnitude longer than currently possible. Hence, one must resort to
simpler, coarse-grained models in order  to study the kinetics of interaction
between a PDZ domain and its target peptide.

Neglecting the atomic detail clearly has both advantages and drawbacks. On the
one hand,  the reduction in the number of degrees of freedom and the simplified
force-field of interaction  allow to  explore longer times scales, thus
making a complete equilibrium description of the binding kinetics possible.
On the other hand, the chemical specificities
can be reincorporated in the model only at a qualitative level.

In this paper, we employ the G$\bar{\rm o}$ strategy~\cite{cecilia}, a native-centric 
scheme, coarse-grained at the residue
level to simulate the unbinding dynamics of a peptide from the third PDZ domain 
of PSD-95 (henceforth referred to as PDZ3). Such simplified description has already 
proved successful in the characterization of the 
role of the native state topology and of its dynamics in protein recognition and binding 
mechanisms~\cite{fly-casting,HIVCecconi}. 
Our simplified simulation scheme allows us to draw a clear, though 
approximated, picture of the dissociation kinetics. 

The escape of the peptide from
the binding groove on the PDZ surface takes place over  a free-energy landscape
that strongly depends on the temperature. At physiologically relevant
temperatures, less than 10\% lower than the unfolding temperature
(measured to be about 320K for the second PDZ domain of
PTP-BL~\cite{Gianni:2005}), the free energy difference between the bound state
and the barrier turns out to be rather small ($0.5$ kcal/mol) because of the 
nearly complete compensation of the enthalpic and entropic components. 
As a consequence, unbinding at physiological temperatures is eminently a diffusive, 
rather than thermally activated, process and the unbinding
rate is orders of magnitude smaller than an Arrhenius-like extrapolation 
from low temperatures would suggest.

\section{Methods}

The PDZ3 domain (PDB codes 1BFE and 1BE9 without and with bound peptide respectively)
as resolved by X-ray crystallography~\cite{Doyle:1996}
is 110 residue long but we truncated the chain from 
Arg309 to Ser393
because the final loop does not participate to the binding mechanism and its
large flexibility may hinder some signals specifically involved in the 
binding.  

We adopt the model proposed by Clementi {\it et al.}~\cite{cecilia}. 
where successive beads along the chain are connected by stiff harmonic springs,
mimicking the peptide bond and maintaining the chain connectivity,
\begin{equation}
V_{\rm bb}(\vec{r}_{i,i+1}) = \frac{1}{2}K \epsilon (r_{i,i+1}-R_{i,i+1})^2
\end{equation}
with  stiffness $K = 1000/d_0^2$, where  $d_0 = 3.8$ \AA \ is the
mean equilibrium distance of two consecutive residues along the chain, and 
$\epsilon=0.84$ kcal/mol (see Results) sets the energy scale. 
Here, $r_{ij}$ and $R_{ij}$ indicate the distance between residues
$i$ and $j$ in a generic conformation and in the native structure (1BFE and 1BE9), respectively.

In line with native-centric schemes,
non-bonded (nb) interactions between non consecutive $\alpha$-carbons 
are modelled with Lennard-Jones 12-10 potentials if the atoms are 
in contact in the native state according to a given interaction cutoff $R_{c}$ 
and with purely repulsive interactions 
otherwise 
\begin{equation}
V_{\rm nb}(\vec{r}_{ij}) = 
\begin{cases}
\epsilon \left[5 \left( \frac{\ds R_{ij}}{\ds r_{ij}} \right)^{12} -
6 \left (\frac{\ds R_{ij}}{\ds r_{ij}} \right)^{10} \right] & R_{ij} \leq R_{c} \\
\frac{\ds 2\epsilon}{\ds 3} 
\bigg( \frac{\ds \sigma}{\ds r_{ij}} \bigg)^{12} & R_{ij} > R_{c}
\end{cases}
\end{equation}
The parameters of the non-bonded interactions are fixed as $R_c=6.5$~\AA \ and  $\sigma= 4.5$~\AA.
The force field is completed by the angular interactions 
\begin{eqnarray}
V_{\rm ang} = \sum_{i=2}^{N-1} \frac{1}{2}k_{\theta}(\theta_{i} - \theta_{i}^0)^{2} +
\nonumber \\
\sum_{i=3}^{N-2} k_{\phi}^{(1)}[1 - \cos(\phi_{i} - \phi_{i}^{0})] +
k_{\phi}^{(3)}[1 - \cos3(\phi_{i} - \phi_{i}^{0})]
\nonumber \\
\end{eqnarray}
where $\theta_{i}$ is the bending angle identified by the three consecutive
C$_{\alpha}$'s $i-1$, $i$, $i+1$ and  $\phi_{i}$  is the dihedral angle defined
by the two adjacent planes formed by four consecutive 
C$_{\alpha}$'s at $i-2$, $i-1$,
$i$, $i+1$. The  superscript $0$ identifies quantities referring to the native
conformation. The force field parameters are proportional to the energy scale
$\epsilon$  so that $k_{\theta} =
20\epsilon$, $k_{\phi}^{(1)} = \epsilon$, $k_{\phi}^{(3)} =0.5\epsilon$
and one time unit corresponds to about $3$ ps, when considering an average 
aminoacid mass of 110 Da.

We have performed fixed-temperature Molecular Dynamics simulations within the 
isokinetic scheme~\cite{Isokin}, which provides a correct sampling of the configuration space. 
We have then applied the multiple histogram technique~\cite{SF} 
to estimate thermodynamic observables such as the internal energy, the specific 
heat
\begin{equation}
C_v = \frac{\langle V^2 \rangle - \langle V \rangle^2 }{T^2}
\end{equation}
and  the structural similarity $Q$ of a given conformation with the 
native structure. The latter parameter is defined as
\begin{equation}
Q = \frac{\sum_{i>j} \Theta(R_c - R_{ij}) \Theta(R_c - r_{ij})}
{\sum_{i>j} \Theta(R_c - R_{ij})}
\label{eq:overlap}
\end{equation}
$\Theta(u)$ being the unitary step function. 
$Q$ represents the fraction of native contacts present in a 
given conformation.
The multiple histogram technique allows to construct the free energy profiles 
$G(Q) = -k_{B} T \log P(Q)$ as functions of $Q$, which plays the role 
of a reaction coordinate.

In order to obtain the free energy profiles of unbinding, we employed the 
umbrella sampling technique~\cite{Umbrella}. In particular, we restrained  the 
distance between the peptide and the PDZ3 centers  of mass 
$\rho=|\vec{R}_{\rm pdz} - \vec{R}_{\rm pept}|$ 
to a given range of  values $D_k$ $(k=1,2,\dots,N_{\rm samp})$ via the 
harmonic umbrella potential
$$
V_{u}(\rho) = \frac{K_q}{2}(D_{k} - \rho)^2
$$
with $K_q = 0.84$ kcal/mol/\AA$^2$. 
The multiple histogram technique allows then a de-biasing at a given 
temperature by matching all the  histograms for the center-mass 
distance collected around the different  sampling values. 
By varying the reference 
temperature we can obtain different free energy curves.

The statistics of the unbinding times have been collected through Langevin dynamics.
The unbinding time is defined here as the time
at which the bond distances of all the 13 contacts linking  the peptide to the PDZ3
exceed a given threshold for the first time. We fixed such thresholds at 1.5 times
the corresponding values in the native conformation. However, small variations
of this threshold did not result in major changes of the exit times statistics.

\section{Results}

\begin{figure}
\includegraphics[clip=true,width=8.9truecm]{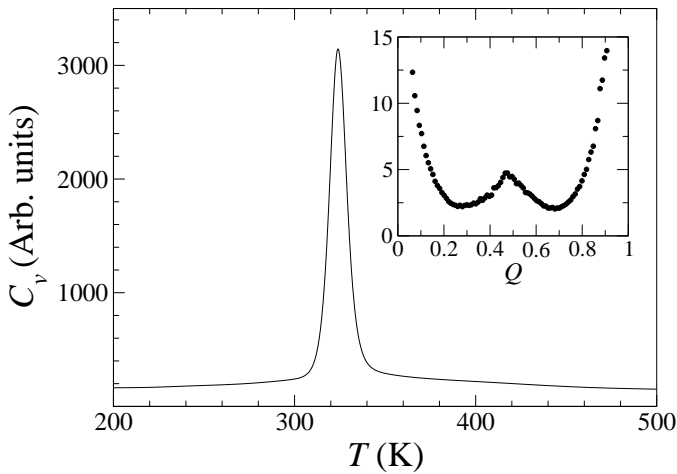}
\caption{Specific heat curve of G\={o}-model
PDZ domain folding/unfolding. The inset shows the double well feature of 
the free energy profile (kcal/mol) as a function of the fraction of native 
contacts $Q$ at the folding temperature.}
\label{fig:param}
\end{figure}

The energy scale $\epsilon$ can be fixed by comparing the folding temperature 
$T_f$ of the model, in units of $\epsilon/k_B$, with experiments.
For this purpose we studied the PDZ-domain specific heat $C_v$ as a function of temperature.
The thermogram in Fig.~\ref{fig:param} allows to identify the unfolding 
transition at $T_f \approx 0.76$.
From available data~\cite{Gianni:2005}, the folding temperature of PDZ3 is 
close to 323 K (50$^{\rm \, o}$ C), which allows to set the model 
energy scale to the value $\epsilon = 0.84$ kcal/mol. 
The free energy profile as a function of the fraction of native contacts $Q$
(Fig.~\ref{fig:param}, inset) is typical of clean two-state transitions,
in agreement with experiments~\cite{Gianni:2005}, where
folding was found to be affected at most by a high energy poorly populated intermediate.

A further check of the chosen energy scale
may be obtained from a comparison with 
the experimental stability $\Delta G_0 = 6-6.5$ kcal/mol of PDZ3 at 
298 K \cite{Gianni:2005}. 
From our simulations we can estimate $G_{0}$ as the energy difference
between the minima in the native and denaturated basins.
At $T=298$ K we find $\Delta G_0 = 5$ kcal/mol, in good
agreement with experiments.

\begin{figure}
\includegraphics[clip=true,keepaspectratio,width=8.2truecm]{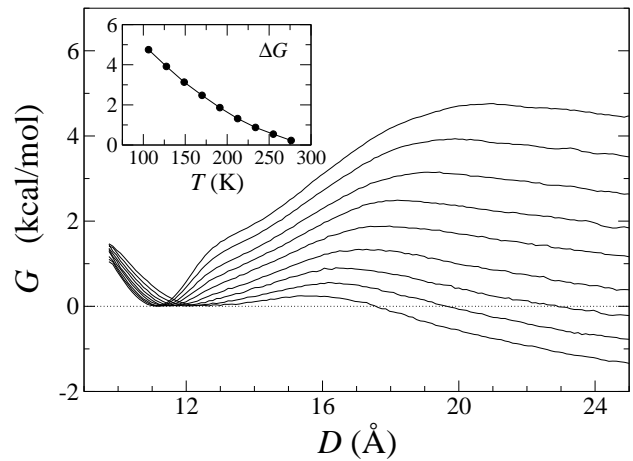}
\caption{Free energy profile along the reaction coordinate at different 
temperatures below the unfolding of the complex PDZ3-Peptide.
From top to bottom, the curves refer to temperatures from $T=85$ K  up 
to $T=255$ K in steps  $\Delta T = 21$ K. The inset shows the temperature 
dependence of the free energy barrier (kcal/mol).}
\label{fig:free_en}
\end{figure}

It is our aim to investigate the unbinding dynamics of a small peptide 
from the PDZ3 domain. Hence, both the  unbinding and unfolding 
transitions have to be  carefully located in temperature. 
The free-energy profiles as a function of the distance between the centers of 
mass of the protein and of the peptide, chosen as reaction coordinate for the unbinding, 
are reported in Fig.~\ref{fig:free_en} for  several temperatures
lower than that of unbinding. 
At low temperatures, the curves are characterized by a well defined minimum 
corresponding to the bound state and by a steep free energy barrier that 
has to be overcome in order for the complex 
to thermally dissociate. Remarkably, the barrier turns out to decrease rapidly  
at increasing temperatures, marking an almost complete compensation of 
the enthalpic barrier by the increase of conformational entropy. 
The behaviour of the dissociation free 
energy $\Delta G(T) = G_{\rm barr}(T) - G_{\rm min}(T)$  
with the temperature is shown in the inset of 
Fig.~\ref{fig:free_en}. As expected, $\Delta G$ is a decreasing function 
of $T$, vanishing at $T_{\rm diss} =  306$ K, 
which locates the spontaneous unbinding  17$^{\rm \, o}$ C below the unfolding 
temperature. 

\subsection{The unbinding dynamics}  

\begin{figure}[t!]
\includegraphics[clip=true,keepaspectratio,width=8.cm]
{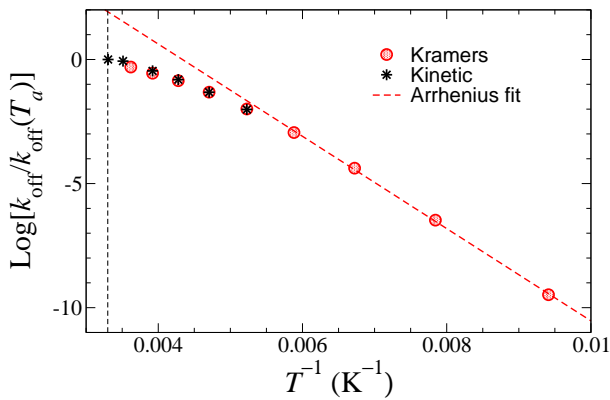}
\caption{Arrhenius plot comparing the dissociation rates estimated from 
the free energy profiles of Fig.~\ref{fig:free_en} through Eq.~\eqref{eq:Kramers}
and the rates computed as inverse average times from direct unbinding simulations. 
Since pre-factors are unknown by construction, the latter have been rescaled by an 
appropriate temperature-independent factor to match the values obtained from Kramers' theory. 
All rates are normalized by the kinetic rate at room temperature $T_{a}=300$ K.
The thick dashed line is the result of an Arrhenius fit with the expression 
$k_{\rm off} = a \exp(-\Delta/k_{B}T)$, which gives a free energy barrier height of 
$\Delta \simeq 7$ kcal/mol
. The vertical dashed line marks the room temperature.} 
\label{fig:Arrhenius}
\end{figure}

The data shown in Fig.~\ref{fig:free_en} contain the relevant information 
on the unbinding dynamics. In fact, it is possible to compute the 
dissociation rates by straight integration 
of the numerical free energy profiles according to Kramers' theory 
\cite{Kramers},
\begin{equation}
k_{\rm off}(T) = \frac{\displaystyle k_{B}T}
{\displaystyle \gamma \int_{x_{0}}^{x_b} \mbox{e}^{\frac{G(x)}{k_B T}} \, dx
\int_{0}^{x} \mbox{e}^{-\frac{G(x^{\, \prime})}{k_B T}}\, dx^{\, \prime}}
\label{eq:Kramers}
\end{equation}  
where $x_0$ and $x_b$ are the abscissas of the minimum and of the barrier 
maximum, respectively.  
The temperature curve of the dissociation rates 
as calculated from formula~\eqref{eq:Kramers} is plotted in 
Fig.~\ref{fig:Arrhenius}.
Our results clearly show that the simple Arrhenius description 
$k_{\rm off} \propto \exp(-\Delta G/k_{B}T)$ only holds in the low temperature 
region,
where the free energy barrier is high enough to justify the familiar 
treatment of a thermally activated unbinding. 
At higher temperatures
the dissociation process slows down considerably with respect to the 
extrapolated low-temperature prediction.
 
This effect is due to the gradual switch from a barrier-limited, thermally 
activated process at low
temperatures to a diffusion-limited process as the unbinding temperature 
is approached. Indeed, using (\ref{eq:Kramers}) in the limit case where
there is a perfectly flat free-energy landscape between $x_0$ and $x_b$ 
(and a reflecting wall in $x_0$), hence no barrier, 
$k_{\rm off} = 2 (k_{B} T/\gamma)/ (x_b-x_0)^2$, which is the typical rate for
particles diffusing from $x_0$, 
with diffusion constant $k_{B} T/\gamma$, and absorbed at $x_b$. 
The diffusion-limited $k_{\rm off}$ depends only slightly on temperature when 
represented on an Arrhenius plot (Fig.~\ref{fig:Arrhenius}).  

However, the straightforward application of Kramers' theory at high 
temperatures may not be legitimate due to the strong reduction of the free 
energy barriers, that
makes the hypotheses of local equilibrium in the free energy minimum and high 
barrier questionable.
     
\begin{figure}[t!]
\includegraphics[clip=true,keepaspectratio,width=8.5truecm]{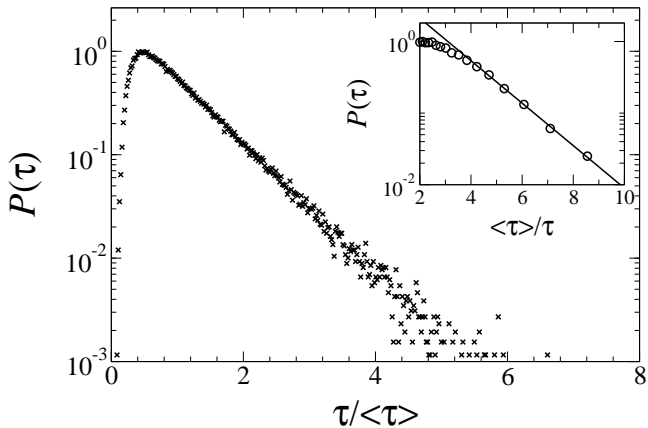}
\caption{Histogram of the kinetic unbinding times at $T = 300$ K (symbols).
The solid line in the inset is a fit of the early times region with the
first passage time distribution of a diffusive process 
$P(\tau) = a \exp(-\tau_{0}/\tau)$.}
\label{fig:Ptau300K}
\end{figure}
In order to check the validity of the Kramers' prediction at high temperatures, 
we carried out kinetics simulations of unbinding to directly estimate the 
dissociation rate at a temperature $T$. 
We observed the time evolution of the PDZ3/peptide complex until a 
spontaneous unbinding event was recorded. 
By doing so, we were able to measure very accurately the 
lifetimes $\tau$ of the molecular complex, the 
temperature-dependent average $\langle \tau \rangle$ and distributions 
$P(\tau)$. The inverse of $\langle \tau \rangle$ provides a measure of the 
kinetic dissociation rate: $k_{\rm off} = 1/\langle \tau \rangle$.
The results of this procedure are shown in Figure~\ref{fig:Arrhenius} 
together with 
the rates estimated from Kramers' formula (\ref{eq:Kramers}).  
The agreement between  unbinding simulations and the analytical estimates 
is remarkable at all temperatures where we were able to collect enough 
statistics for the computation of the average inverse lifetime of the complex. 
In fact, dissociation events become exponentially
rarer at low temperatures, thus requiring unrealistic simulation 
times in order to collect satisfactory statistics. 
These results provide a direct confirmation of
the validity of Eq.~\eqref{eq:Kramers} in the high temperature region.

That the unbinding process is diffusion-limited at high temperatures
is nicely confirmed by the analysis of
the the histogram of dissociation times. 
In Fig.~\ref{fig:Ptau300K}
we show one of such distributions computed at $T=300$ K. 
At long times $P(\tau)$ decreases exponentially, implying that the 
slow unbinding events obey the statistics of the waiting 
times of a Poisson-like process, as should be expected for
a simple Arrhenius picture of thermal activation over a barrier.
However, it can be clearly appreciated that fast events are under-represented 
with respect to what the Arrhenius picture would predict. 
In particular (see inset of Fig.~\ref{fig:Ptau300K}),
we find that the left tail of the distribution follows the law 
$P(\tau) \propto \exp(-\tau_{0}/\tau)$, which indeed characterizes the 
distribution of first passage times in a diffusive process.

\begin{figure}[t!]
\includegraphics[keepaspectratio,width=8.5truecm,clip]{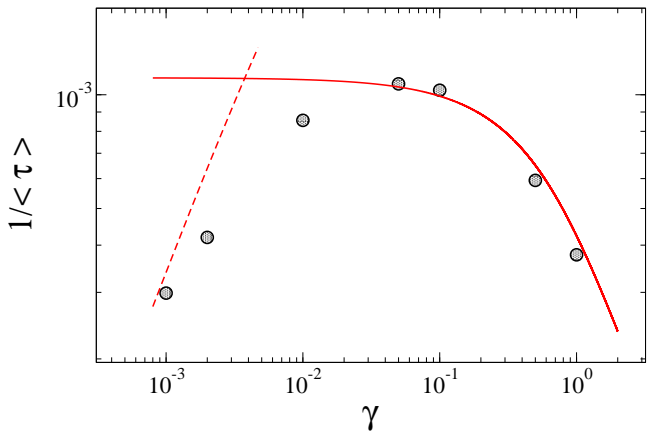}
\caption{Average kinetic rate at the intermediate temperature 
$T = 234$ K as a function of the friction 
$\gamma$ used in the Langevin simulation of the unbinding kinetics. 
The solid and dashed lines are the predictions of Kramers' 
theory~\eqref{eq:rateod} calculated with the frequencies and
free energy barrier estimated from the corresponding simulated free energy 
profiles of Fig.~\ref{fig:free_en}. Note that no fitting is performed here.} 
\label{fig:taugamma}
\end{figure}

\begin{figure}[t!]
\includegraphics[clip=true,keepaspectratio,width=8.5truecm]{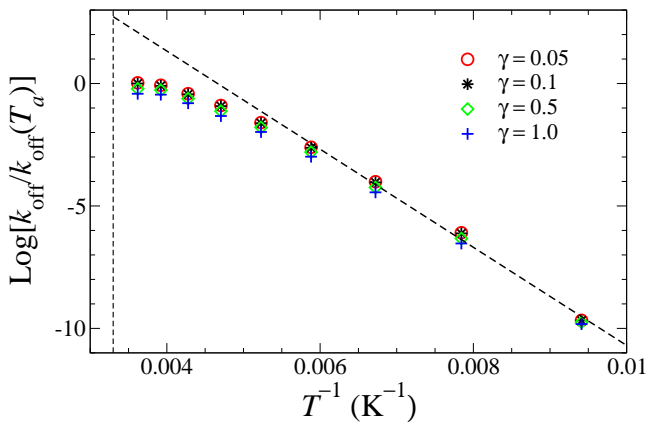}
\caption{Arrhenius plot comparing the dissociation rates estimated for
different values of $\gamma$ from 
the over--damped formula in eq.~\eqref{eq:rateod} with the 
temperature--dependent parameters $\omega_{0}$, $\omega_{b}$
and $\Delta G$ estimated from the simulated free energy profiles.
All rates are normalized by the kinetic rate at room temperature $T_{a}=300$ K
for $\gamma = 0.05.$
The vertical dashed line marks the room temperature.} 
\label{fig:Arrhenius_gamma}
\end{figure}

As a final  methodological remark, we wish to comment on using the 
Langevin scheme for
the unbinding simulations. 
It is known that the kinetics simulated with Langevin MD is strongly dependent 
on the friction  coefficient $\gamma$. 
We thus performed different runs 
at different values of $\gamma$ in order
to check the dependence of the unbinding kinetics on the friction. 
The results of this
analysis are reported in Fig.~\ref{fig:taugamma}, where we plot the 
average peptide dissociation time $\langle \tau \rangle$ at an intermediate
temperature $T=234$ K 
as a function of $\gamma$. 
We explored both the under-damped and the over-damped regimes, 
and compared the numerics with the asymptotic limits of Kramers' theory of 
barrier crossing~\cite{melnikov:1018}
\begin{equation}
\label{eq:rateod} 
k_{\rm off} \approx  \\ 
\begin{cases} 
\frac{\ds \omega_{0} e^{-\frac{\Delta G}{k_B T}} }{\ds 2 \pi} \left[ 
                         \frac{\ds \gamma S_{1}}{\ds T} - 0.82 \left( \frac{\ds \gamma S_{1}}{\ds T}\right)^{3/2} 
                         \right]  & \gamma S_{1} \ll T \\
\frac{\ds \omega_{0} e^{-\frac{\Delta G}{k_B T}} }{\ds 2 \pi} \left[ 
                         \sqrt{\frac{\ds \gamma^2}{\ds 4 \omega_{b}^2}+1} - \frac{\ds \gamma}{\ds 2\omega_{b}} 
                         \right]   & \gamma S_{1} \gg T 
\end{cases}
\end{equation}
where $\omega_{0}$  and $\omega_{b}$ are the frequencies at the bottom and saddle points of the free energy profile
and 
\[
S_{1} = 2 \int_{x_{1}}^{x_{b}} \sqrt{-2m [G(x)-G(x_{b})]} \, dx \quad ,
\]
$x_{1}$ being the left--hand side turning point, i.e. the solution of the 
equation $G(x_{1})=G(x_{b})$
with $x_{1}< x_{0}$.
We find that our data indeed converge toward  Kramers' estimates in both 
the under--damped and the over--damped regimes, thus showing that the 
computation of 
$k_{\rm off}$ as $1/\langle \tau \rangle$ can be meaningfully 
extrapolated to different  values of $\gamma$ at different temperatures.
In particular, we can make use of the theoretical prediction for the 
over--damped regime, seen to be valid 
for values of $\gamma$ approximately greater than $0.1$, for testing whether 
the results of Fig.~\ref{fig:Arrhenius}
still hold for other values of the damping parameter when using frequencies and
free energy barriers computed numerically. 
We see from Fig.~\ref{fig:Arrhenius_gamma} that the essential features of 
the Arrhenius plot are preserved. 
Importantly, the room temperature value of the dissociation rate  
remains still largely overestimated 
if extrapolated from the low--temperature values even at larger damping 
$\gamma$.
 
The scenario outlined above is consistent with other MD 
studies employing the Langevin scheme to investigate the kinetics of 
folding~\cite{thirumalai}.


\subsection{Unbinding dynamics of individual contacts}
A more detailed description of the unbinding process
can be obtained by considering separately the kinetics of the $13$
contacts that keep the peptide bound to the PDZ in the native state. 
Each of them will be characterized by a different breaking time 
$\langle \tau_c \rangle$, that can be estimated as the average 
over many dissociation events in a typical unbinding simulation.  
The ranking between such times
provides an average  picture of the order at which contacts break down 
during the gradual detachment of the peptide from the PDZ domain. 

\begin{figure}[t!]
\includegraphics[clip=true,keepaspectratio,width=9.0truecm]{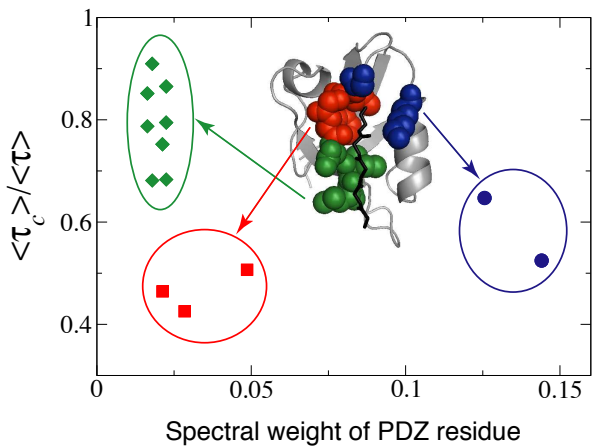}
\caption{Average breaking times of individual contacts versus the spectral
weight on the subset of functional modes of the corresponding residue in 
the PDZ domain. The breaking times are normalized to the average 
unbinding time of the peptide as a whole.} 
\label{fig:contact}
\end{figure}

It is thus interesting to investigate whether
there exists a correlation between the unbinding event of each 
PDZ/peptide bond and the dynamics of the residues from the PDZ domain 
involved in the same link.
A relation between the binding dynamics and a specific spatial pattern 
characterizing a reduced subset of low-frequency normal modes
has been recently suggested by De Los Rios {\it et al.}~\cite{immortal}. 
In the present case, we find four modes among the eigenstates 
of the Hessian of our force field that match the pattern 
identified in Ref.~\onlinecite{immortal}, namely the set $\mathcal{S}=\{7,8,9,14\}$.
We thus introduce an indicator measuring the spectral weight 
corresponding to that set of modes at each site  
\begin{equation}
   \label{e:specw}
     w_i  = \sum_{k \in \mathcal{S}} \sum_{\alpha=x,y,z} [\xi_{i,\alpha}^k]^2
   \end{equation}
where $\xi_{i,\alpha}^k$ is the $\alpha$ ($\alpha=\{x,y,z\}$) 
component of the $k$-th normal mode at site $i$.	    
In figure~\ref{fig:contact} we plot $\langle \tau_c \rangle$ vs. $w_i$ 
for the 13 binding contacts, where $i$ indicates the residue of the PDZ domain 
participating to the contact $c$ with the peptide.

The data show to be clustered into three sets. 
The right outermost cluster contains the two contacts that have both a short lifetime and the larger
spectral weight. In terms of the latter indicator, these contacts are characterized by 
the fourth and fifth largest values among all residues. 
This strongly suggests that functional normal modes are likely to contribute substantially to the 
loosening of the PDZ/peptide bond and its  eventual rupture. 
The left bottom cluster involves the contacts with small 
$\langle \tau_c \rangle$ but 
   whose local fluctuations are poorly captured by the functional modes. 
   Taken together, these first two clusters include the residues forming
   the hydrophobic pocket: the second group comprises the residues at the 
   bottom of the pocket whereas the first one includes two residues flanking the
   entry of the binding pocket. 
   Finally, the last contacts to break up involve the less 
conserved residues that confer specificity to the different PDZ domains.

Overall, these results outline the following picture of a typical 
unbinding event. The first contacts to loosen up are those involving
residues from the hydrophobic binding pocket, part of which are involved
in the {\em breathing} pattern of the cleft as highlighted from normal
mode analysis. This observation provides a nice example of the interpretation 
of a specific normal mode pattern as {\em precursory} feature of a much larger
amplitude motion.
The contacts that break up the last are those from the surface region of the 
binding pocket, the ones associated with binding specificity and thus the less 
conserved ones. Since these residues are obviously more important as their chemical details
are concerned, it is not surprising that a dynamical investigation based on
a purely topological model does not assign them a key role in the 
unbinding kinetics.

\subsection{Conclusions and Discussion}
The central result of our analysis of the unbinding of peptides from
the PDZ3 domain is that the bound state is stabilized over a physiologically 
relevant temperature interval.

Counterintuitively, this happens as a direct consequence of the strong 
temperature--dependent quenching of the free energy barrier
in the proximity of the functional region of temperatures.
As a matter of fact, such slow down of the dissociation reflects the gradual change in 
nature of the fastest events from thermally activated to diffusive. 
Indeed, as the free energy barrier gets substantially reduced at 
increasing temperatures due to the entropy-enthalpy compensation
typical in the proximity of the unfolding transition,
Kramers' formula~\eqref{eq:Kramers} predicts
a diffusion dominated unbinding process, and kinetic 
simulations confirm such picture.
Overall, diffusion slows down the process of two orders of 
magnitude with respect to the prediction of the Arrhenius 
formula extrapolated from the low-temperature region.

The spectral analysis of the breaking time of the contacts involved in the
binding helps shedding further light on the unbinding process,
that on average first involves the escape from the hydrophobic binding pocket 
and subsequently from the surface region where residues conferring specificity
are usually located.  From the time history of all the contacts that keep the 
peptide bound to the domain until they rupture we obtained their 
average lifetimes, that correlate with a picture emerged form NMA 
analysis. 
This suggests a non trivial coupling between collective dynamics
of the molecule and binding mechanism even at physiological temperatures. 
This confirms that Normal Mode Analysis may provide a valuable method for 
detecting the protein chain fluctuations that represent the precursory
events to unbinding.    

Our results rest on the reliable calculation of the free-energy profile 
of unbinding, which is a quantity defined at (quasi-)equilibrium, 
and on the precise determination of the escape-time histograms, 
which implies a complete description of the metastable state of 
the process, again a (quasi-)equilibrium quantity. This is consistent with 
the observation that experimental unbinding rates range from $1$ to $10^{-3}$ 
s$^{-1}$, time-scales long enough for nanometric dynamical molecules such as 
PDZ domains to fully explore their phase-space and consequently to be amenable 
to a (quasi-)equilibrium description. 

Unfortunately, present all-atom MD with realistic force-fields can access at 
most few hundreds nanoseconds of the dynamics, and are thus likely to miss events relevant 
on the second or longer time-scales.
Employing a native centric backbone representation of the protein with 
simplified force field (G\={o}-like model) on the one hand allowed us a thorough
investigation of the equilibrium and stationary properties of the system,
and on the other hand emphasized the role of the native state geometry in the 
binding mechanism of the PDZ3 domain. 
Given the overall high structural similarity of the PDZ family,
our evidence for a diffusion-limited unbinding process likely applies 
to most if not all of them.

Although necessary to access the relevant time-scales, neglecting the chemical 
details of the binding cleft and of the peptide comes at a price: it is 
impossible to precisely rank the affinities of different peptides for the same
domain and of different domains for the same peptide. At the same time,
the promiscuity of PDZ binding implies that some degree of generalized stickiness, 
captured by the simplified force-field used here, is present.
Moreover, we believe that our results represent a reading frame for future 
unbinding simulations that will become possible when
algorithms and computational resources will be powerful enough
to access the experimental time-scales of the process while taking
into account the full chemical detail.

At the same time, we stress that our prediction of diffusive vs. thermally activated unbinding 
is surely amenable to experimental verification.


\end{document}